\documentclass[prb,twocolumn,superscriptaddress,showpacs,longbibliography]{revtex4-1}
\usepackage{graphicx}
\usepackage{epsfig}
\usepackage{amssymb,amsmath,amsfonts,hyperref}
\usepackage{wasysym}
\usepackage{latexsym}
\usepackage{eucal}
\usepackage{verbatim}
\usepackage[normalem]{ulem}
\usepackage{color}

\newcommand{\be}{\begin{eqnarray}}
  \newcommand{\ee}{\end{eqnarray}}

\begin{document}
\title{Singular ferromagnetic susceptibility of the transverse-field Ising antiferromagnet
on the triangular lattice}
\author{Sounak Biswas}
\affiliation{\small{Tata Institute of Fundamental Research, 1 Homi Bhabha Road, Mumbai 400005, India}}
\author{Kedar Damle}
\affiliation{\small{Tata Institute of Fundamental Research, 1 Homi Bhabha Road, Mumbai 400005, India}}
\begin{abstract}
  A transverse magnetic field $\Gamma$ is known to induce antiferromagnetic three-sublattice order of
  the Ising spins $\sigma^z$ in the triangular lattice Ising antiferromagnet at low enough temperature. This low-temperature order is known to melt on heating in a two-step manner, with
  a power-law ordered intermediate temperature phase characterized by power-law
  correlations at the three-sublattice wavevector ${\bf Q}$: $\langle \sigma^z(\vec{R}) \sigma^z(0)\rangle \sim \cos({\mathbf Q}\cdot \vec{R}) /|\vec{R}|^{\eta(T)}$ with the temperature-dependent power-law exponent $\eta(T) \in (1/9,1/4)$.
  Here, we use a newly developed quantum cluster algorithm to study the {\em ferromagnetic} easy-axis susceptibility $\chi_{u}(L)$ of an $L \times L$ sample in this power-law ordered phase. Our numerical results are consistent
with a recent prediction of a  singular $L$ dependence $\chi_{u}(L)\sim L^{2- 9 \eta}$ when $\eta(T)$ is in the range $(1/9,2/9)$. This finite-size
  result implies, via standard scaling arguments, that the ferromagnetic susceptibility $\chi_{u}(B)$ to a uniform field $B$ along the easy axis is singular at intermediate
temperatures in the small $B$ limit, $\chi_{u}(B) \sim  |B|^{-\frac{4 - 18 \eta}{4-9\eta}}$ for $\eta(T) \in (1/9, 2/9)$, although there is no ferromagnetic long-range order in the low temperature state.

\end{abstract}
\vskip2pc
\maketitle
\section{Introduction}
The transverse field Ising antiferromagnet on the triangular lattice, with Hamiltonian,
\begin{equation}
  H_{\rm Ising} = J_{1}\sum_{\langle \vec{R} \vec{R}'\rangle }\sigma^z_{\vec{R}} \sigma^z_{\vec{R}'} -\Gamma \sum_{\vec{R}}\sigma^x_{\vec{R}}  -B \sum_{\vec{R}} \sigma^z_{\vec{R}}\; ,
  \label{Hising}
\end{equation}
where $\vec{\sigma}_{\vec{R}}$ are Pauli matrices representing $S=1/2$ moments on sites $\vec{R}$ of the triangular lattice, $\langle \vec{R} \vec{R}'\rangle$ denote the nearest neighbour links of the triangular lattice,  $J_1>0$ is the antiferromagnetic exchange among easy-axis components of the $S=1/2$ moments (a factor of $\frac{1}{4}$, appropriate for
$S=1/2$ moments, has been absorbed in the definition of $J_1$), and $B$ and $\Gamma$ are components of the external magnetic field along the easy axis $\hat{z}$  and transverse direction $\hat{x}$ respectively (a factor of $\frac{g\mu_B}{2}$, appropriate for $S=1/2$ moments, has been absorbed in the definition of these field components), provides perhaps the simplest example of a quantum ``order-by-disorder''~\cite{Villain,Moessner_Sondhi} effect, whereby a classical spin liquid develops long-range magnetic order upon the introduction of terms in the Hamiltonian that induce quantum fluctuations.

When $\Gamma=0$, the zero temperature classical Ising antiferromagnet at $B=0$
has a macroscopic degeneracy of minimum exchange-energy configurations on the triangular lattice. These are in correspondence with all dimer covers of the dual honeycomb lattice,
 implying that the entropy-density remains nonzero in this classical zero temperature limit.\cite{Wannier,Houtappel} At non-zero temperature, thermal fluctuations of $\sigma^{z}$ allow for defects that take the system out of the minimum exchange-energy dimer subspace.
The Ising spins remain in  a paramagnetic state all the way down to $T=0$,\cite{Wannier,Houtappel}  albeit with a diverging correlation length\cite{Stephenson} at the three-sublattice wavevector ${\bf Q}$. This provides a simple example of classical spin
liquid behaviour, with the $T=0$ limit characterized  by power-law spin correlations
at the three-sublattice wavevector ${\bf Q}$.

A transverse field $\Gamma$ that couples to $\sigma^x$ induces quantum fluctuations of
the Ising spins $\sigma^z$, and would ordinarily be expected to further reduce any residual ordering tendency of the Ising spins. However, in reality, these quantum fluctuations immediately stabilize a ground-state
with long-range three-sublattice order of $\sigma^z$ for any nonzero $\Gamma$. In contrast to the {\em ferrimagnetic} three-sublattice order exhibited by the classical Ising antiferromagnet with ferromagnetic further neighbour couplings\cite{Landau}, the $\Gamma >0$ ground state is characterized by {\em antiferromagnetic} three-sublattice order,\cite{Isakov_Moessner} {\em i.e.}, the modulation of $\langle \sigma^z\rangle$ at wavevector ${\mathbf Q}$
is not accompanied by any net ferromagnetic moment. At $T=0$, this three-sublattice ordered persists up to a critical value $\Gamma_{c}\approx 1.7$\cite{Isakov_Moessner} (in units of $J_1$), beyond which the system becomes a quantum paramagnet in which
the spins are polarized in the $\hat{x}$ direction.\cite{Moessner_Sondhi_Chandra,Moessner_Sondhi,Isakov_Moessner}  When the system is heated to nonzero temperatures
above this three-sublattice ordered ground state, the three-sublattice order
melts via an intermediate-temperature phase characterized by power-law order: $\langle \sigma^z(\vec{R}) \sigma^z(0)\rangle \sim \cos({\mathbf Q}\cdot \vec{R}) /|\vec{R}|^{\eta(T)}$ for $T \in (T_{1},T_{2})$,  with a temperature-dependent power-law exponent $\eta(T)$
that is expected\cite{Jose_Kadanoff_Kirkpatrick_Nelson} to increase from $\eta(T_1)=1/9$ to $\eta(T_2)=1/4$.\cite{Isakov_Moessner} 

A recent field-theoretical analysis\cite{Damle} predicts that the {\em ferromagnetic easy-axis susceptibility} $\chi_{u}(B)$ to the uniform longitudinal field $B$ along the easy-axis diverges at small $B$ in 
a large portion of such power-law ordered phases associated with the two-step melting
of three-sublattice order in frustrated easy-axis antiferromagnets with triangular lattice symmetry: $\chi_{u}(B) \sim |B|^{-\frac{4 - 18 \eta}{4-9\eta}}$ for $\eta(T) \in (1/9,2/9)$.
For the specific case of the transverse field  Ising antiferromagnet on the triangular lattice, this is a rather counter-intuitive prediction: 
The Ising spins in Eq.~\eqref{Hising} have no ferromagnetic couplings, and ferromagnetic correlations remain short-ranged in the low-temperature phase with long-range three-sublattice
ordered phase. Yet, the prediction is for the uniform easy-axis susceptibility to start diverging once this three-sublattice order melts partially due to thermal fluctuations. 

Our goal here is to test this general prediction using the test-bed provided by the transverse
field Ising antiferromagnet (Eq.~\eqref{Hising}) on the triangular lattice. In order to do this, we need to obtain an accurate characterization of the long-distance form of the correlations of 
the easy-axis magnetization density as well as correlations of the three-sublattice
order parameter for this model. These can be used to obtain
the finite-size easy-axis susceptibility $\chi_{u}(L)$ of the Ising spins, as well as the
value of $\eta(T)$ for a range of temperatures in the power-law ordered phase.
If the easy-axis susceptibility is indeed singular as predicted, then standard finite-size scaling arguments imply that $\chi_{u}(L) \sim L^{2-9 \eta}$
for $\eta(T) \in (1/9,2/9)$ in the power-law ordered phase. In this paper, we test this form of the prediction using
a newly developed quantum-cluster algorithm\cite{Biswas_Rakala_Damle} that provides an efficient tool
for performing Quantum Monte Carlo simulations of frustrated transverse field
Ising models within the Stochastic Series Expansion\cite{Melko,Sandvik_PRE} framework.

The rest of this paper is organized as follows: In Section.~\ref{PhasesandTransitions}, we discuss the antiferromagnetic nature of the three-sublattice  order induced by the transverse field and contrast it with the ferrimagnetic three-sublattice ordered phase established by additional ferromagnetic couplings. 
We also review the standard Landau theory framework used for describing this kind of
long-range order, and use it to discuss the possible theoretical scenarios for the phase transition between these two phases.
In Section.~\ref{Methods} we provide a brief sketch of the actual
computational method used to obtain our numerical results. In  Section.~\ref{Results}, we 
summarize our results for the uniform magnetization density as well as the
three-sublattice order parameter and compare them with the field-theoretical
predictions alluded to earlier.

\section{Phases and transitions}
\label{PhasesandTransitions}
The \emph{antiferromagnetic} (with no net easy-axis magnetic moment) three-sublattice
order exhibited by the $\Gamma > 0$ ground state of $H_{\rm Ising}$ can be thought
of in terms of the following useful caricature: Ising spins on one spontaneously chosen
sublattice (out of the three sublattices corresponding to the natural tripartite decomposition of the triangular lattice) freeze into the $\lvert \sigma^{x}=+1\rangle$ state. Equivalently,
one may think of them as fluctuating freely between the $\lvert \sigma^{z}=+1\rangle$ and $\lvert \sigma^{x}=-1\rangle$ states due to the effects of quantum fluctuations. On the other two sublattices of the triangular
lattice, the system orders antiferromagnetically, with spins on one sublattice pointing up
along the $\hat{z}$ axis, and spins on the other sublattice pointing down. This is also
the picture for the long-range ordered phase that persists up to the lower-critical
temperature $T_1(\Gamma)$ that marks the onset of the power-law ordered intermediate
phase associated with the two-step melting of three-sublattice order. 

On incorporating an additional next-neighbour ferromagnetic coupling $J_{2}<0$, the antiferromagnetic three-sublattice order of the low-temperature phase gives way to \emph{ferrimagnetic} (with net easy axis moment) three-sublattice order beyond a
non-zero threshold value $J_{2c}$.\cite{Biswas_Rakala_Damle} This is because
the classical ($\Gamma=0$) model with $J_{2} < 0$ is known to develop ferrimagnetic
three-sublattice order beyond a $T=0$ threshold at which $\sigma^z$ have power-law
correlators $\langle \sigma^z(\vec{R})\sigma^z(0) \rangle \sim \cos({\mathbf Q} \cdot \vec{R})/|R|^{\eta}$ with $\eta=1/9$.\cite{Nienhuis_Hilhorst_Blotte}
This ferrimagnetic three-sublattice order can be understood in terms of
the following caricature: The system spontaneously chooses one sublattice on which the
spins all point along the $+\hat{z}$ direction ($-\hat{z}$ direction), while the spins
on the other two sublattices all point along the $-\hat{z}$ direction ($+\hat{z}$ direction). 

With this picture of the low temperature phases in mind, we focus our
attention on the uniform easy axis magnetization $m$ and the complex three-sublattice order parameter $\psi$, defined as
\begin{align}
  \label{orderparameters}
  &m=\frac{1}{L^{2}}\sum_{\vec{R}}\sigma^{z}_{\vec{R}} \\
  &\psi=\frac{1}{L^{2}}\sum_{\vec{R}}\sigma^{z}_{\vec{R}}\exp(i\mathbf{Q} \cdot \vec{R}) 
\end{align}
where $\mathbf{Q}$
is the three-sublattice ordering wave vector ($(2\pi/3,2\pi/3)$ in the standard basis) and $\vec{R}$ represents the coordinates
of triangular lattice sites. In the standard Landau-Ginzburg approach\cite{Domany_Schick_Walker_Griffiths,Domany_Schick,Alexander} to thermal
(nonzero temperature) phase transitions involving such three-sublattice ordered states, the physics of three-sublattice ordering is represented in terms of a classical order parameter field $\psi_{\mathrm{cl}}$, which may be identified with the static (Matsubara frequency $\omega_n = 0$) part of the $\psi$ operator defined above:
\begin{equation}
  \psi_{\mathrm{cl}}=\frac{1}{\beta}\int_0^\beta d\tau \psi(\tau)
\label{OPDEFN}
\end{equation}
Here, we used the usual notation for the imaginary-time analog of Heisenberg operators, $\mathcal{O}(\tau)=e^{\tau H_{\rm{TFIM}}}\mathcal{O}e^{-\tau H_{\mathrm{TFIM}}}$, corresponding to any Schr\"{o}dinger operator $\mathcal{O}$.

In this Landau-Ginzburg framework, the free energy is written as an integral over
a coarse-grained free-energy density $\mathcal{F}(\psi_{\mathrm{cl}})$ that admits
an expansion in powers and gradients of a coarse-grained order-parameter field $\psi_{{\mathrm{cl}}}(\vec{r})$ which may be thought of as a local version of the order parameter
defined in Eq.~\ref{OPDEFN}. Keeping various
low-order terms consistent with the action of various symmetries of the microscopic
Hamiltonian, one writes: 
\begin{align}
  \mathcal{F}(\psi_{\mathrm{cl}})=\kappa \lvert \nabla \psi_{\mathrm{cl}} \rvert^2 +  r\lvert \psi_{\mathrm{cl}}\rvert^{2}+u_4\lvert \psi_{\mathrm{cl}}\rvert^{4} \nonumber\\
+ u_6\lvert \psi_{\mathrm{cl}}\rvert^{6} + \lambda_{6}\lvert \psi_{\mathrm{cl}}\rvert^{6}\cos(6\theta)  \nonumber \\ 
  +\lambda_{12}\lvert \psi_{\mathrm {cl}}\rvert^{12}\cos(12\theta)+\ldots
  \label{LGW}
\end{align}
where, $\theta(\vec{r})$ is the phase of the complex order parameter field $\psi_{cl}(\vec{r})$. 
As usual, one assumes that the coefficients of various terms in this phenomenological free-energy are smooth functions of microscopic parameters. In this approach, three-sublattice ordering corresponds to $r<0$. The sign of $\lambda_{6}$  determines the nature of three-sublattice ordering : $\lambda_{6}>0$ favors antiferromagnetic ordering with the phase $\theta$ pinned at $(2n+1)\pi/6$ ($n = 0,1\dots 5$), while $\lambda_{6}<0$ favors ferrimagnetic ordering with the phase pinned at $(2n)\pi/6$ ($n=0,1\dots 5$). The $\lambda_{12}$ term is not expected to be important except when $\lambda_6$ is driven to the vicinity of zero by the competition between further-neighbour ferromagnetic couplings ( in the microscopic Hamiltonian ) that favour ferrimagnetic three-sublattice ordering, and other effects (such as quantum fluctuations induced by a transverse field) that favour antiferromagnetic three-sublattice ordering.  

If fluctuations of $\theta$, the phase of the order parameter, play a dominant role in driving
the transition to a paramagnetic high-temperature state, one expects a phase-only
description to capture the long-wavelength properties near such a transition. In other words, one then expects that $|\psi_{\mathrm{cl}}|$, the amplitude of the order parameter, remains
nonzero near the transition (corresponding to $r<0$), and the physics of the transition
is controlled by the interplay between the effective phase-stiffness $\kappa |\psi_{\mathrm{cl}}|^2$ and the six-fold anisotropy 
$\lambda_6$. This gives rise to the expectation\cite{Domany_Schick_Walker_Griffiths,Domany_Schick,Alexander} of critical behaviour in the universality
class of the six-state clock model\cite{Jose_Kadanoff_Kirkpatrick_Nelson,Challa_Landau,Cardy} of statistical mechanics.

As is well-known, two-dimensional six-state clock models represent an
unusual example of a system which can display a variety of critical behaviours, each
of which is a generic possibility that can be realized for a range of microscopic parameters.\cite{Cardy} Of particular interest in the present context is the possibility of a two-step
melting transition, whereby the low-temperature phase with long-range order in 
$\exp(i\theta)$ is separated from a high-temperature paramagnetic phase by an intermediate
phase with power-law order in $\exp(i\theta)$. As is well known, this
power-law ordered phase is controlled by a line of Gaussian fixed points\cite{Jose_Kadanoff_Kirkpatrick_Nelson} with
effective free-energy density $\frac{1}{4\pi g}\int d^{2}r(\nabla \theta)^{2}$.
For $g\in(1/9,1/4)$, the six-fold anisotropy $\lambda_6$ and the vorticity
in the $xy$ field $\theta$ are both irrelevant perturbations of this fixed-point free-energy
density, which controls the long-wavelength behaviour of order parameter correlations
in the intermediate power-law ordered phase. The continuously varying power-law
exponent $\eta(T)$ for order parameter correlations, which serves as a ``universal coordinate'' that locates a given microscopic system within this power-law ordered phase, is set by the coupling constant $g$ via the relation $\eta(T) = g(T)$.

The Landau-Ginzburg theory also sheds light on the nature of the low temperature transition between the two kinds of three-sublattice ordered phases, modeled by $\lambda_6$ going through zero smoothly and changing
sign. Since both phases have long-range three-sublattice order, fluctuations of $\lvert \psi_{\mathrm{cl}}\rvert$ may again be neglected in the vicinity of this transition. With the amplitude $|\psi_{\mathrm{cl}}|$ remaining essentially constant across this transition, the physics
of the transition is again controlled by the phase $\theta$ of the three-sublattice order parameter. Minimizing the free-energy density $\mathcal{F}$ yields a spatially uniform configuration with a particular optimal value $\theta^{*}$ for this phase variable.  When $\lambda_{12}<0$, $\theta^{*}$ takes on the
values $(2n+1)\pi/6$ ($(2n)\pi/6$) with $n=0,1\dots 5$ when $\lambda_{6}>0$ ($\lambda_{6}<0$ ). When $\lambda_6=0$, all values $\theta^*=m\pi/6$ $(m = 0,1\dots 11$) minimize the free-energy. Clearly, this corresponds to a first-order transition between
ferrimagnetic and antiferromagnetic three-sublattice ordered states, with both kinds of
three-sublattice order coexisting at the transition point. 

If, on the other hand, $\lambda_{12}>0$, we obtain
\begin{equation}
  \theta^{*}=\begin{cases}
    \frac{2n\pi}{6} & \text{if } \lambda_{6} <-4\lambda_{12}\lvert\psi_{\rm cl}\rvert^{6}\\
    \frac{2n\pi}{6}+ \frac{1}{6}\arccos(-\lambda_{6}/4\lambda_{12}\lvert\psi_{\rm cl}\rvert^{6}) & \text{if } \lvert \lambda_{6}\rvert <4\lambda_{12}\lvert\psi_{\rm cl}\rvert^{6}\\
    \frac{(2n+1)\pi}{6} & \text{if } \lambda_{6} >4\lambda_{12}\lvert\psi_{\rm cl}\rvert^{6}\\
  \end{cases}
\end{equation}
where  $n=0,1\dots 5$ represents the six-fold degeneracy of the minima in each case. 
In this case, as $\lvert \lambda_{6} \rvert$ becomes small and $\lambda_6$ goes through zero, $\theta^{*}$ switches continuously from the antiferromagnetic phase to the ferrimagnetic phase via an intermediate mixed phase that is established for $\lvert \lambda_{6} \rvert<4\lambda_{12}\lvert\psi_{\rm cl}\rvert^{6}$. In what follows, we will confront these two
quite different scenarios with data obtained in the vicinity of the transition between
antiferromagnetic and ferrimagnetic three-sublattice order in the low-temperature
state of $H_{\rm Ising}$ with an additional ferromagnetic second-neighbour coupling $J_2$ between the Ising spins.
\begin{figure}[t]
  \includegraphics[width=\columnwidth]{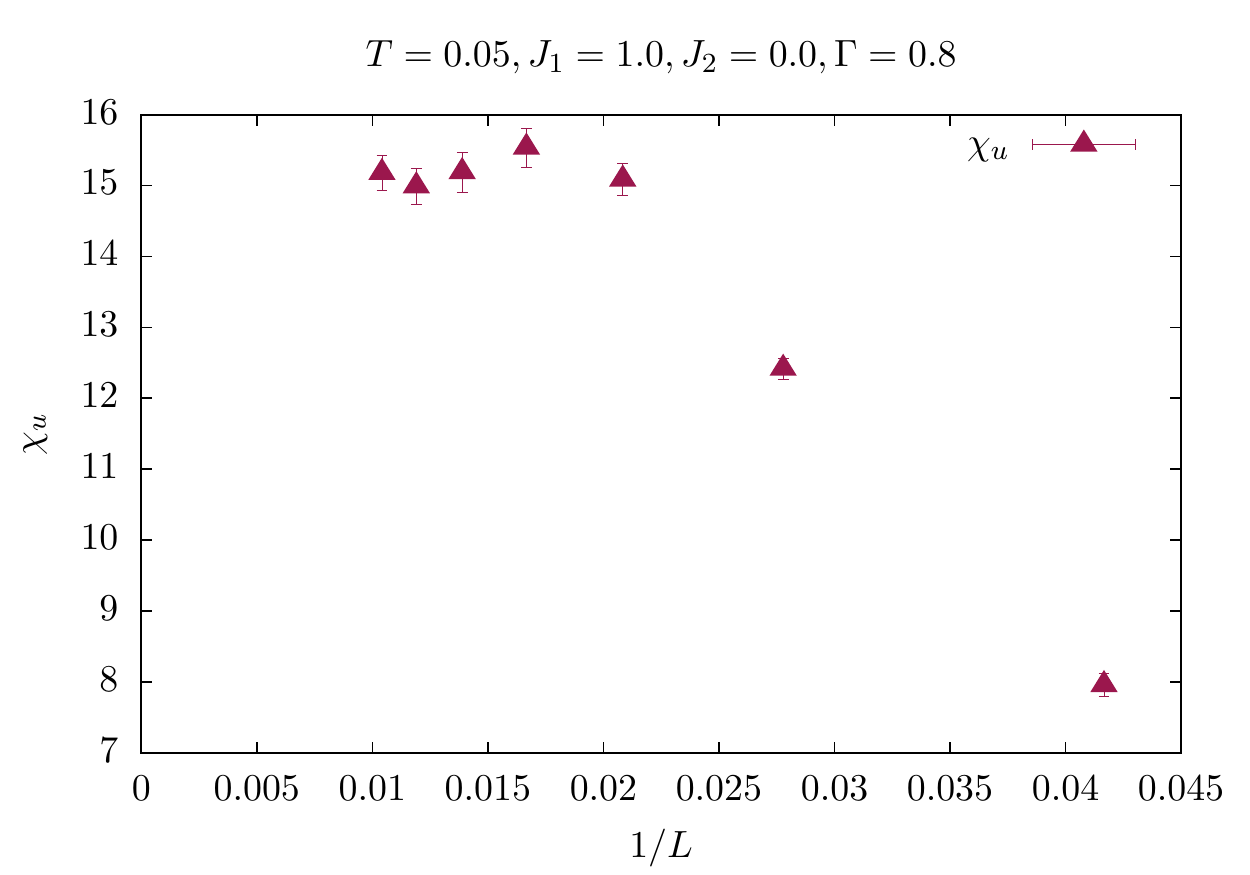}
  \caption{\label{plateau} The uniform easy-axis susceptibility $\chi_{u}$ of $H_{\rm Ising}$ on $L\times L$ triangular lattices, when plotted vs $1/L$ for a sequence of sizes, clearly saturates to a finite value in the limit of large $L$. Note the slow crossover to this thermodynamic limit,
with samples of linear size as large as $L^{*}=40$ not yet in the asymptotic large-$L$ regime. This behaviour demonstrates that the low temperature phase is indeed antiferromagnetic. However, the slow crossover indicates the presence of a proximate phase
with net magnetic moment along the easy-axis, suggesting that $H_{\rm Ising}$ could
be driven into a ferrimagnetic ground state for relatively small values of an additional second-neighbour ferromagnetic coupling $J_2$.  All other
temperature and energy scales are measured in units of $J_1$ which is set to unity.} 
\end{figure}
\begin{figure}[t]
  \includegraphics[width=\columnwidth]{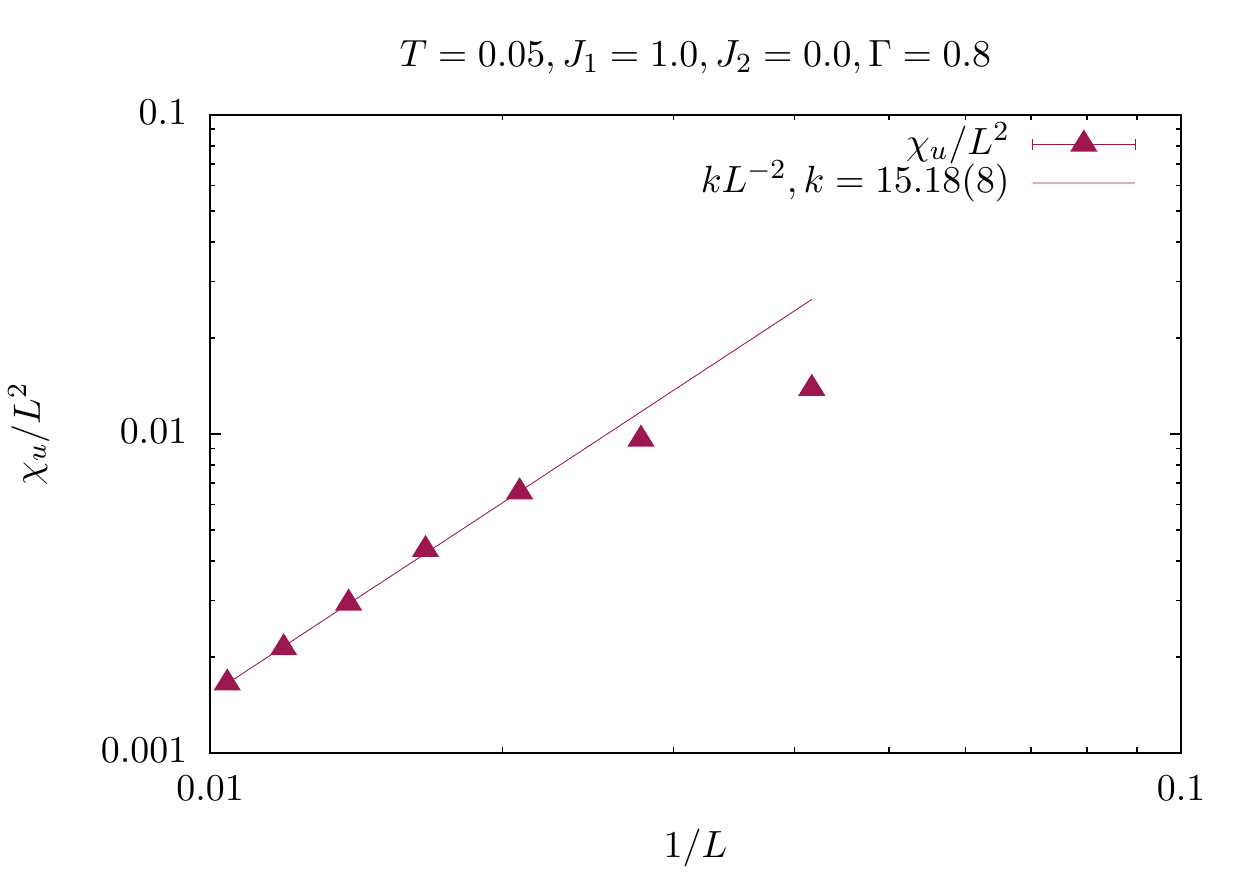}
  \caption{\label{plateaufit} The uniform easy-axis susceptibility of $H_{\rm Ising}$ on $L \times L$ triangular lattices, now scaled
by the number of sites $L^2$, is fit reasonably well to the single parameter form $kL^{-2}$ with $k=15.18(8)$ for the largest four sizes studied here. This analysis also confirms that the low temperature phase of $H_{\rm Ising}$ is indeed antiferromagnetic, {\em i.e.} with no net easy-axis moment. All other
temperature and energy scales are measured in units of $J_1$ which is set to unity.} 
\end{figure}
\begin{figure}[t]
  \includegraphics[width=\columnwidth]{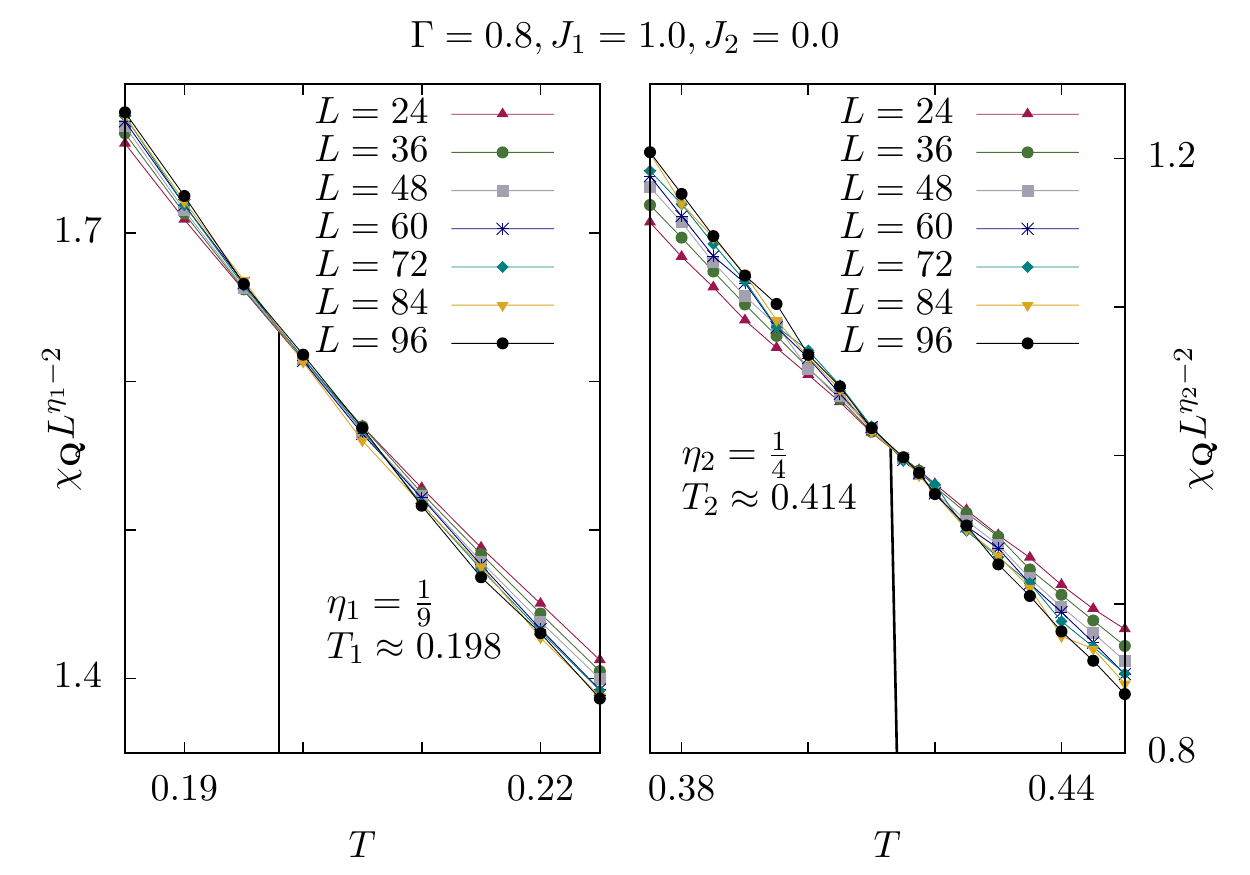}
  \caption{\label{cross1} Lower and upper transition temperatures $T_1$  and $T_2$, which mark the boundaries of the power-law ordered phase associated with the two-step melting of antiferromagnetic three-sublattice order,  are obtained by plotting $\chi_{\mathbf{Q}}L^{\frac{1}{9}-2}$ and $\chi_{\mathbf{Q}}L^{\frac{1}{4}-2}$ versus $T$ for different values of $L$ and identifying
the temperatures at which curves corresponding to different $L$ cross. This
gives $T_{1}=0.198(5)$ and $T_{2}=0.414(5)$ when $\Gamma=0.8$. All other
temperature and energy scales are measured in units of $J_1$ which is set to unity.} 
\end{figure}
\section{Methods}
\label{Methods}

Our numerical work uses  the Stochastic Series Expansion (SSE) framework \cite{Melko,Sandvik_PRE,Sandvik_JPHYSA,Syljuasen_Sandvik,Sandvik_PRB} to compute equilibrium averages $\langle \dots \rangle$ for transverse field Ising models at nonzero temperature. For models with geometric frustration,
which results in a macroscopic degeneracy of minimally frustrated classical configurations
(with minimum Ising-exchange energy), it is important that the computational
method correctly captures the interplay between this macroscopic degeneracy,
and the disordering effects of classical and quantum fluctuations. In the present case, this interplay is expected to be crucial to the establishment of antiferromagnetic three-sublattice order in the low temperature phase,
as well as its two-step melting.\cite{Isakov_Moessner,Moessner_Sondhi_Chandra,Moessner_Sondhi} 

Therefore, to obtain reliable results, we use the recently developed quantum cluster algorithm\cite{Biswas_Rakala_Damle} that works within the SSE framework to provide an
efficient way of sampling the partition function for such frustrated transverse field
Ising models. In this cluster algorithm, which works in the $\sigma^z$ basis, the diagonal Ising exchange part of $H_{\rm{Ising}}$ in Eq.~\eqref{Ising} is written as $\mathcal{H}_{diag}=\sum_{\triangle}\mathcal{H_{\triangle}}$, where $H_{\triangle}$ are operators living on elementary triangular plaquettes $\triangle$. This furnishes the algorithm local information
that enables it to distinguish between minimally 
frustrated plaquettes and fully frustrated plaquettes of higher Ising-exchange energy. The transverse field part of the Hamiltonian is represented as single-site operators as
in the original SSE approach.\cite{Sandvik_PRE} The plaquette representation of $\mathcal{H}_{diag}$ facilitates the construction of ``space-time clusters'' with a broad
distribution of cluster sizes, allowing the algorithm to efficiently sample
the configuration space of SSE operator strings at low temperature.

Using this approach, we study $H_{\rm Ising}$ on $L\times L$ triangular lattice with periodic boundary conditions, with $L$ ranging from $L=24$ to $96$. We compute the static susceptibilities corresponding to the order parameters defined in Eq.~\eqref{orderparameters}. These susceptibilities are defined as
\begin{align}
  &\chi_{u}=\frac{L^{2}}{\beta}\langle\lvert \int_{0}^{\beta} d\tau  m(\tau)\rvert^{2}\rangle \label{chi0}\\
  &\chi_{{\mathbf{Q}}}=\frac{L^{2}}{\beta}\langle\lvert \int_{0}^{\beta} d\tau \psi(\tau)\rvert^{2}\rangle
  \label{chi}
\end{align}
Additionally, we compute the static susceptibility $\chi^{xx}_{{\mathbf{Q}}}$ to a transverse field (along $\hat{x}$) oscillating at wavevector ${\mathbf Q}$, defined as
\begin{equation}
  \chi^{xx}_{{\mathbf{Q}}}=\frac{L^{2}}{\beta}\langle\lvert \int_{0}^{\beta} d\tau \sigma^{x}_{\mathbf{Q}}(\tau)\rvert^{2}\rangle
  \label{chix}
\end{equation}
where $\sigma^{x}_{\mathbf{Q}}$ is given by
\begin{equation}
  \sigma^{x}_{\mathbf{Q}}=\frac{1}{L^{2}}\sum_{\vec{R}}\sigma^{x}_{\vec{R}}\exp(i\mathbf{Q} \cdot \vec{R}) 
\end{equation}

\section{Results}
\label{Results}
We begin by revisiting the phase diagram obtained in previous work\cite{Isakov_Moessner} for the case with no next-nearest neighbour coupling ($J_{2}=0$).
From their results, we note that the low temperature order persists up to
the highest temperature when $\Gamma$ is in the vicinity of $\Gamma = 0.8$.
Therefore, we set the transverse field to this value in most of our work and study
the three-sublattice ordering of the low temperature phase, as well as its
two-step melting.

As expected, we find that the order parameter susceptibility $\chi_{{\mathbf{Q}}}$ scales
with the volume of the system at low enough temperature, confirming the presence of long-range three-sublattice order in the low  temperature phase. Since this is entirely
consistent with earlier results,\cite{Isakov_Moessner} we do not display this explicitly here.
Since our focus in what follows will be an unusual singular behaviour in
the ferromagnetic susceptibility $\chi_{u}$ to a {\em uniform} field along the easy-axis,
we find it useful to first study the same quantity deep in the low-temperature
ordered state. From Fig.~\ref{plateau} and Fig.~\ref{plateaufit}, which display the $L$ dependence of $\chi_{u}$ and $\chi_{u}/L^2$ deep in the low-temperature ordered state, we see that the three-sublattice ordering in the low temperature phase is not accompanied by
any net moment along the easy-axis. 

This confirms earlier results\cite{Isakov_Moessner}
that have identified the antiferromagnetic nature of the three-sublattice ordering
at low temperature. However, the approach to the
thermodynamic limit is seen to involve a slow crossover, suggesting the presence of
a proximate phase with a net easy-axis moment. This is consistent with the fact
that a relatively small value of second-neighbour ferromagnetic exchange $J_2 < 0$ is
sufficient to access a nearby state with ferrimagnetic three-sublattice
ordering at low temperature.\cite{Biswas_Rakala_Damle}

In the power-law ordered phase associated with the two-step melting of three-sublattice order, the static susceptibility $\chi_{\mathbf{Q}}$, defined in Eq.~\eqref{chi} for a finite size $L \times L$ system, is expected to scale as 
\begin{equation}
  \chi_{\mathbf{Q}}\sim L^{2-\eta(T)}
  \label{chiQdecay}
\end{equation}
From the renormalization group picture (summarized in the previous section) of
this power-law ordered phase, it is also clear that $\eta(T)$ ranges from
$\eta(T_1)=1/9$ at the lower phase boundary $T_1(\Gamma)$ of the power-law phase, to $\eta(T_2)=1/4$ at the upper phase boundary $T_2(\Gamma)$.

To locate these upper and lower transition temperatures for $\Gamma=0.8$, we plot $\chi_{\mathbf{Q}}L^{\frac{1}{9}-2}$ and  $\chi_{\mathbf{Q}}L^{\frac{1}{4}-2}$ for various sizes $L$ as a function of temperature and identify the temperature at which curves corresponding to the
different sizes all cross. This is shown in Fig.~\ref{cross1}. The location of transitions obtained in this way are consistent 
with those obtained earlier in Ref.~\onlinecite{Isakov_Moessner}.
\begin{figure}[t]
  \includegraphics[width=\columnwidth]{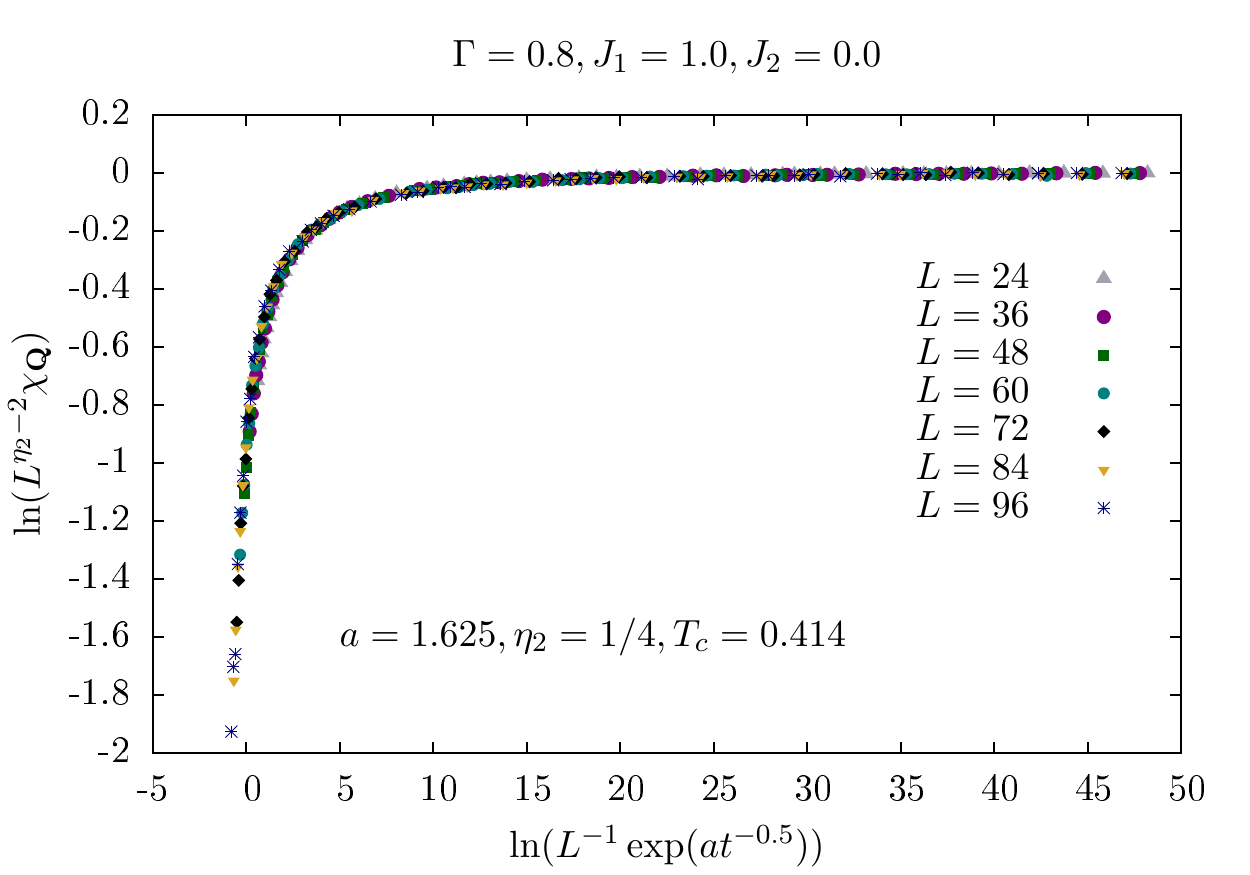}
  \caption{\label{scaling1} Quantum Monte Carlo data for the static susceptibility $\chi_{\mathbf{Q}}$ of $H_{\rm Ising}$ at
wavevector ${\mathbf{Q}} $ on $L\times L$ triangular lattices collapses onto a universal
scaling form when  $\chi_{\mathbf{Q}}(t,L)L^{\frac{1}{4}-2}$ for different $L$ and temperatures $T$ (in the vicinity of the upper transition temperature $T_2$) are plotted as a function of
the scaling variable defined in Eq.~{\protect{\eqref{finitesize}}} in the main text. All other
temperature and energy scales are measured in units of $J_1$ which is set to unity.} 
\end{figure}

\begin{figure}[t]
  \includegraphics[width=\columnwidth]{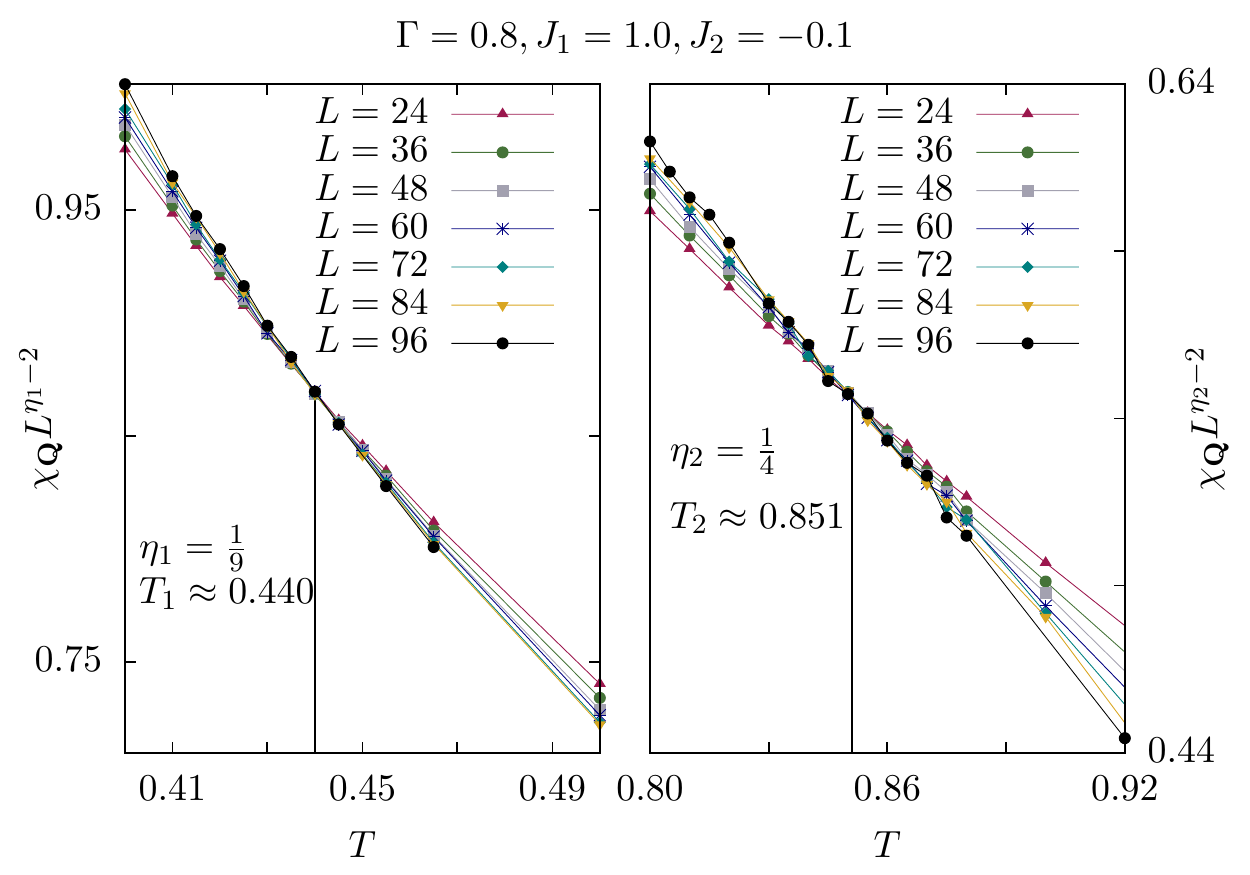}
  \caption{\label{cross2} The ferrimagnetic three-sublattice order that characterizes
the ground state in the presence of a second-neighbour ferromagnetic interaction  $J_{2}=-0.1$ also melts in a two-step manner. Upper and lower transition temperatures $T_1$ and $T_2$, that demarcate the boundaries of the power law ordered phase associated with this two-step
melting, are obtained by plotting $\chi_{\mathbf{Q}}L^{\frac{1}{9}-2}$ and $\chi_{\mathbf{Q}}L^{\frac{1}{4}-2}$ versus $T$ for different values of $L$ and identifying
the temperatures at which curves corresponding to different $L$ cross. This
gives $T_{1}=0.440(6)$ and $T_{2}=0.851(8)$ when $\Gamma=0.8$. All other
temperature and energy scales are measured in units of $J_1$ which is set to unity.} 
\end{figure}
\begin{figure}[t]
  \includegraphics[width=\columnwidth]{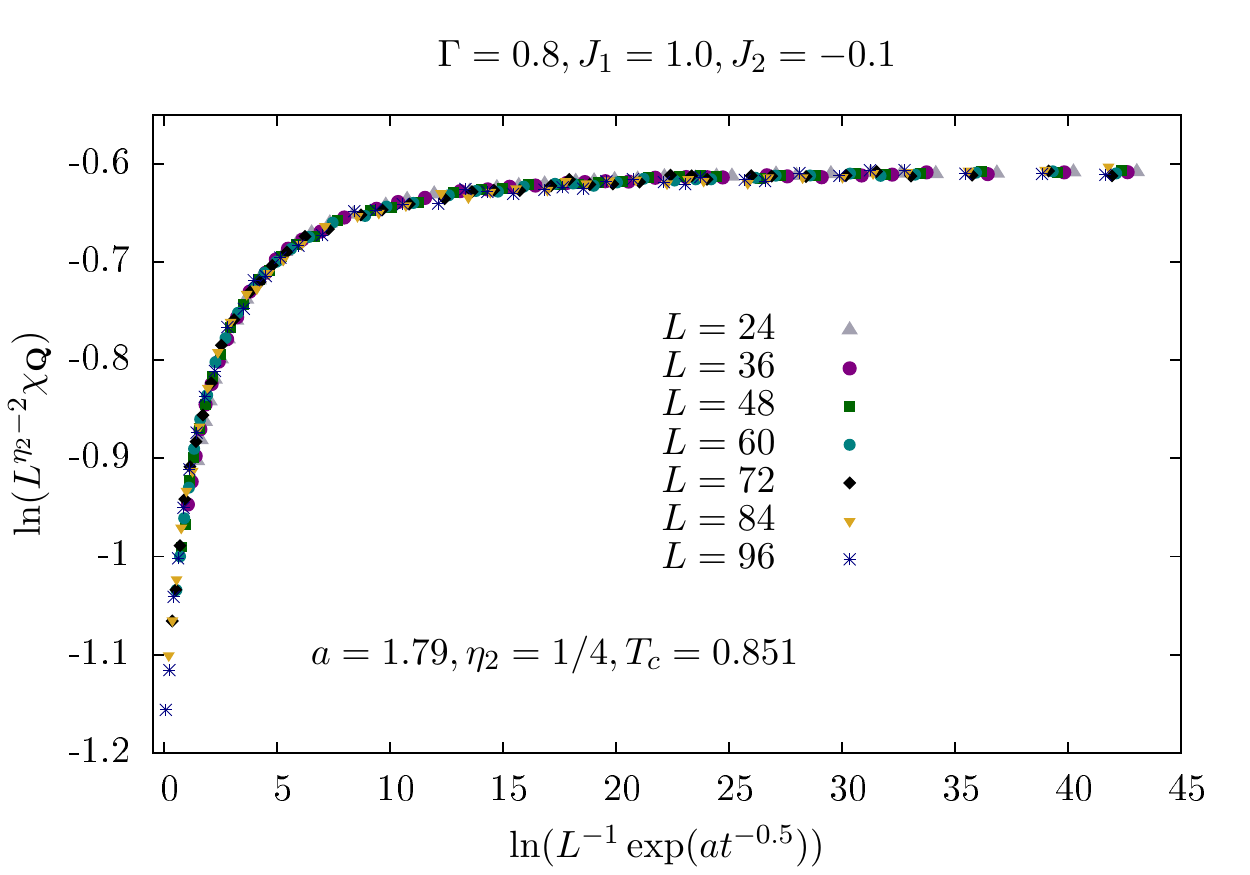}
  \caption{\label{scaling2} Quantum Monte Carlo data for the static susceptibility $\chi_{\mathbf{Q}}$ of $H_{\rm Ising}$ with $J_2=-0.1$ at
wavevector ${\mathbf{Q}} $ on $L\times L$ triangular lattices also collapses onto a universal scaling form when  $\chi_{\mathbf{Q}}(t,L)L^{\frac{1}{4}-2}$ for different $L$ and temperatures $T$ (in the vicinity of the upper transition temperature $T_2$) are plotted as a function of
the scaling variable defined in Eq.~{\protect{\eqref{finitesize}}} in the main text. All other
temperature and energy scales are measured in units of $J_1$ which is set to unity.} 
\end{figure}

Since the upper (lower) transitions out of the power-law ordered phase correspond
to vorticity (six-fold anisotropy) in $\theta$ becoming relevant, we expect
these transitions to be of the Kosterlitz-Thouless (inverted Kosterlitz-Thouless) type.
To confirm that this is indeed the case, we perform fits of our Quantum Monte Carlo data 
in the vicinity of the upper phase boundary to the finite-size scaling form predicted by Kosterlitz-Thouless theory.\cite{Challa_Landau} This scaling form follows from
the following argument: Above $T_2(\Gamma)$, order parameter correlations decay
exponentially, with a correlation length $\xi$ given by\cite{Kosterlitz}
\begin{equation}
  \xi \sim \exp(at^{-1/2}) \; ,
  \label{xi}
\end{equation}
where $t=(T-T_{2})/T_{2} $ is the reduced temperature.
This Kosterlitz-Thouless form of the correlation length, Eq.~\eqref{xi}, in conjunction
with the standard finite size scaling ansatz $\chi_{\mathbf{Q}}(t,L)=L^{2-\eta_2}f(\xi/L)$ gives
the finite-size scaling form\cite{Challa_Landau}
  \begin{equation}
    \chi_{\mathbf{Q}}(t,L)L^{\frac{1}{4}-2}=f(L^{-1}\exp(at^{-1/2})) \ ;,
    \label{finitesize}
  \end{equation}
where we have used $\eta_2=1/4$, and $f$ is the finite-size scaling function that
we expect our data to collapse onto. In practice, we use
$T_{2}$ obtained from Fig.~\ref{cross1}, and attempt a finite-size
scaling collapse with a single adjustable parameter $a$. This is shown in Fig.~\ref{scaling1}.
\begin{figure}[t]
    \includegraphics[width=\columnwidth]{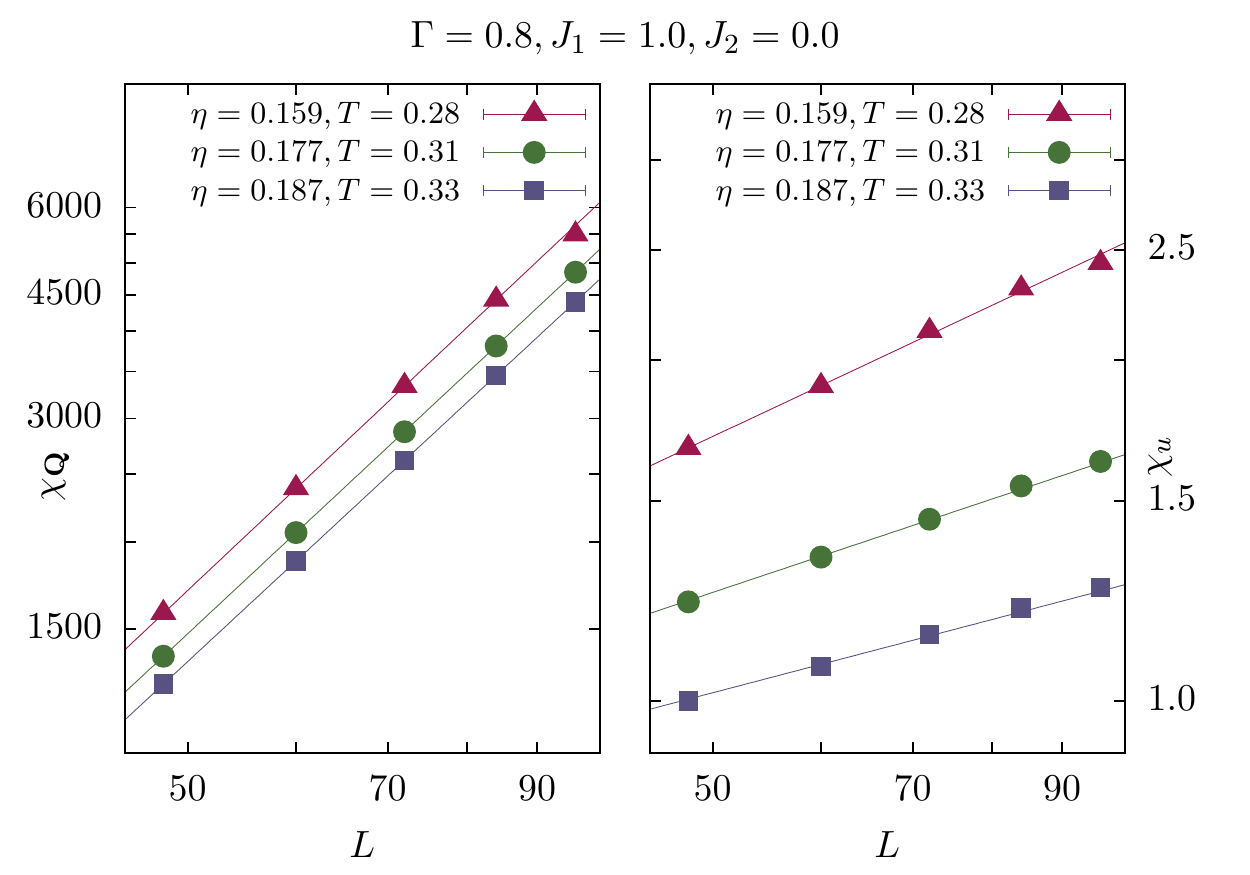}
    \caption{\label{powerlaw1} $\chi_{\mathbf{Q}}$ and $\chi_{u}$ fit rather well
to power-law forms $k_{1}L^{2-\eta}$ and $k_{2}L^{2-9\eta}$ respectively for three different values of temperature in the intermediate power-law ordered phase associated with the melting of antiferromagnetic three-sublattice order when $J_{2}=0.0$, $\Gamma=0.8$. All other
temperature and energy scales are measured in units of $J_1$ which is set to unity.} 
  \end{figure}
  \begin{figure}[t]
    \includegraphics[width=\columnwidth]{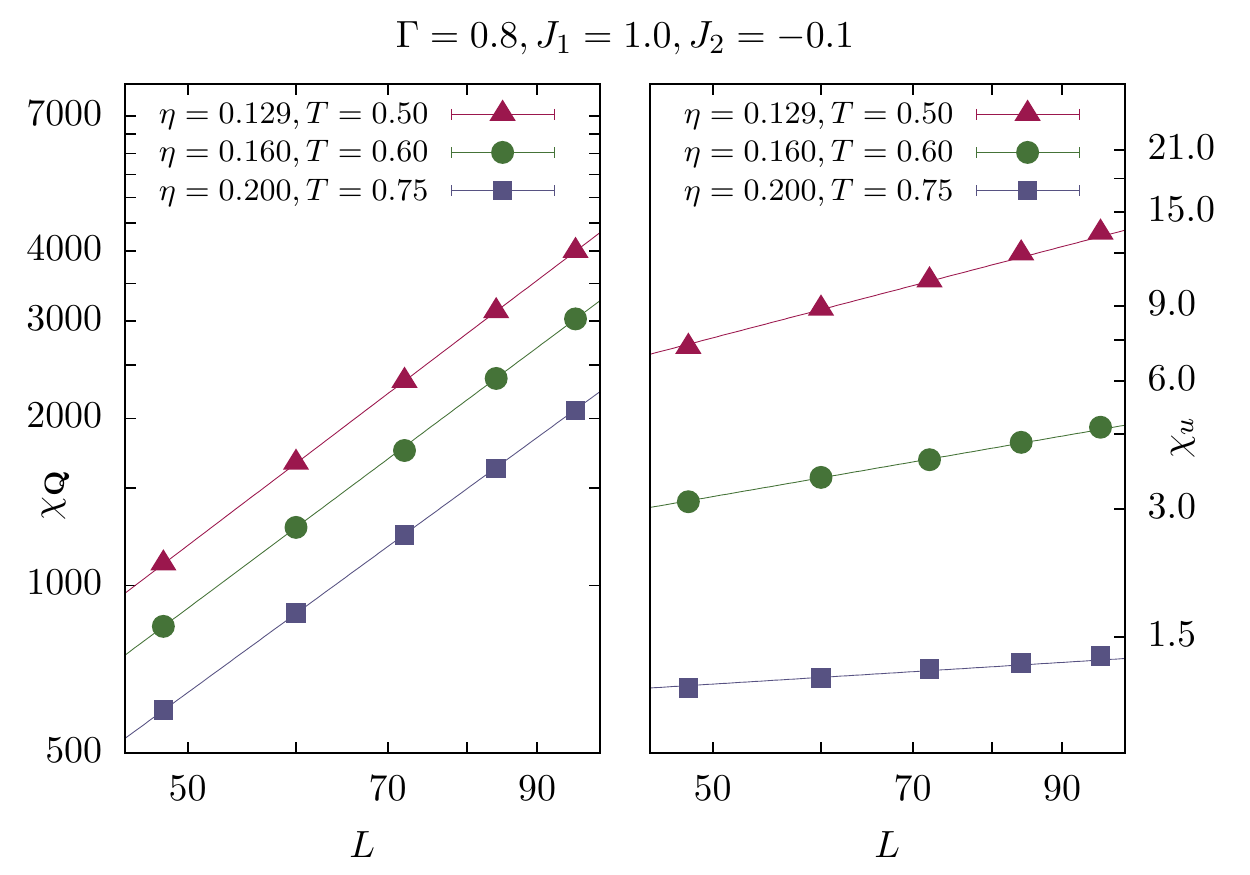}
    \caption{\label{powerlaw2} $\chi_{\mathbf{Q}}$ and $\chi_{u}$ fit rather well
to power-law forms $k_{1}L^{2-\eta}$ and $k_{2}L^{2-9\eta}$ respectively for three different values of temperature in the intermediate power-law ordered phase associated with the melting of ferrimagnetic three-sublattice order when $J_{2}=-0.1$, $\Gamma=0.8$. All other
temperature and energy scales are measured in units of $J_1$ which is set to unity.} 
  \end{figure}
\begin{figure}[t]
    \includegraphics[width=\columnwidth]{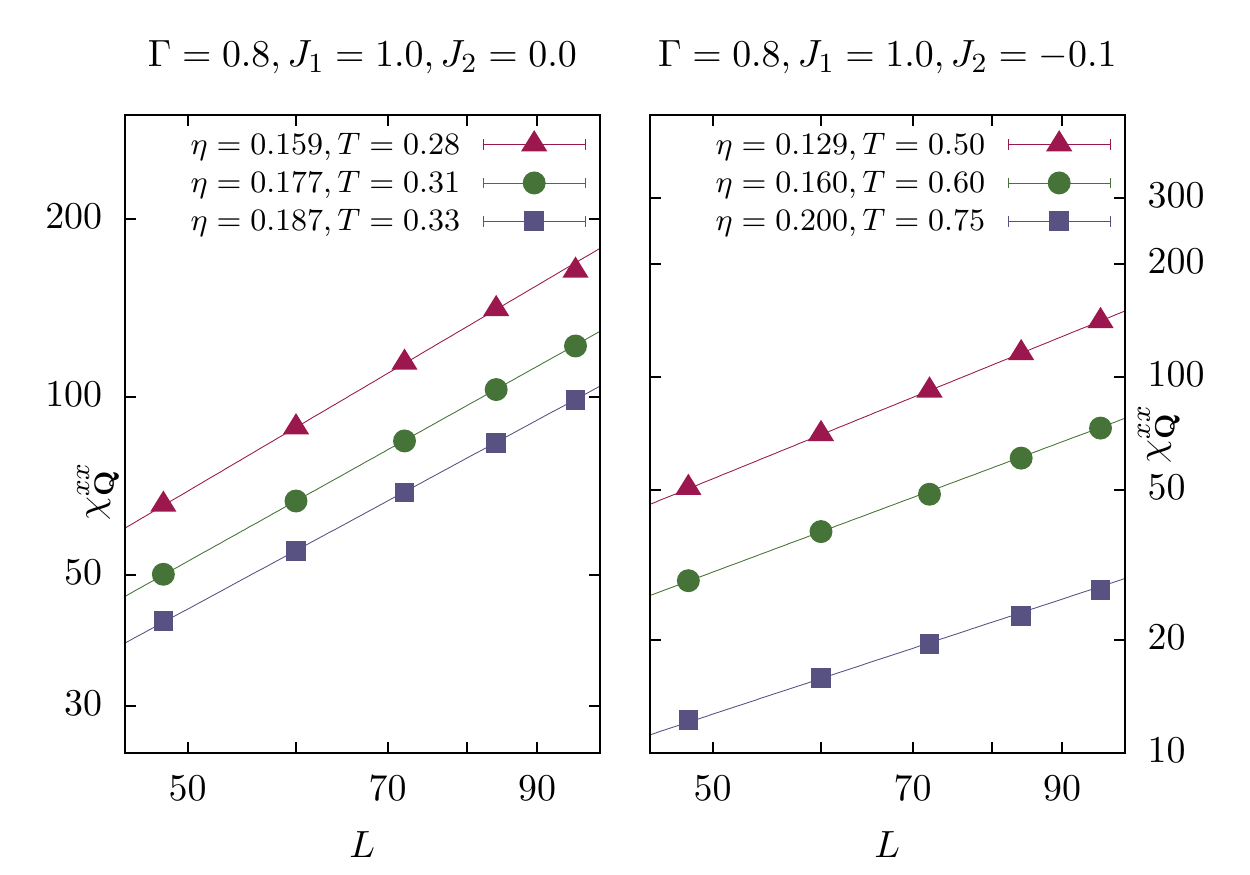}
    \caption{\label{transversepowerlaw} $\chi^{xx}_{\mathbf{Q}}$ fits the power-law
form $k_3 L^{2-4\eta}$ for three different values of temperature in the intermediate power-law ordered phase associated with the melting of antiferromagnetic as well as ferrimagnetic three-sublattice order. All other temperature and 
energy scales are measured in units of $J_1$ which is set to unity.} 
  \end{figure}
 \begin{figure}[t]
    \includegraphics[width=\columnwidth]{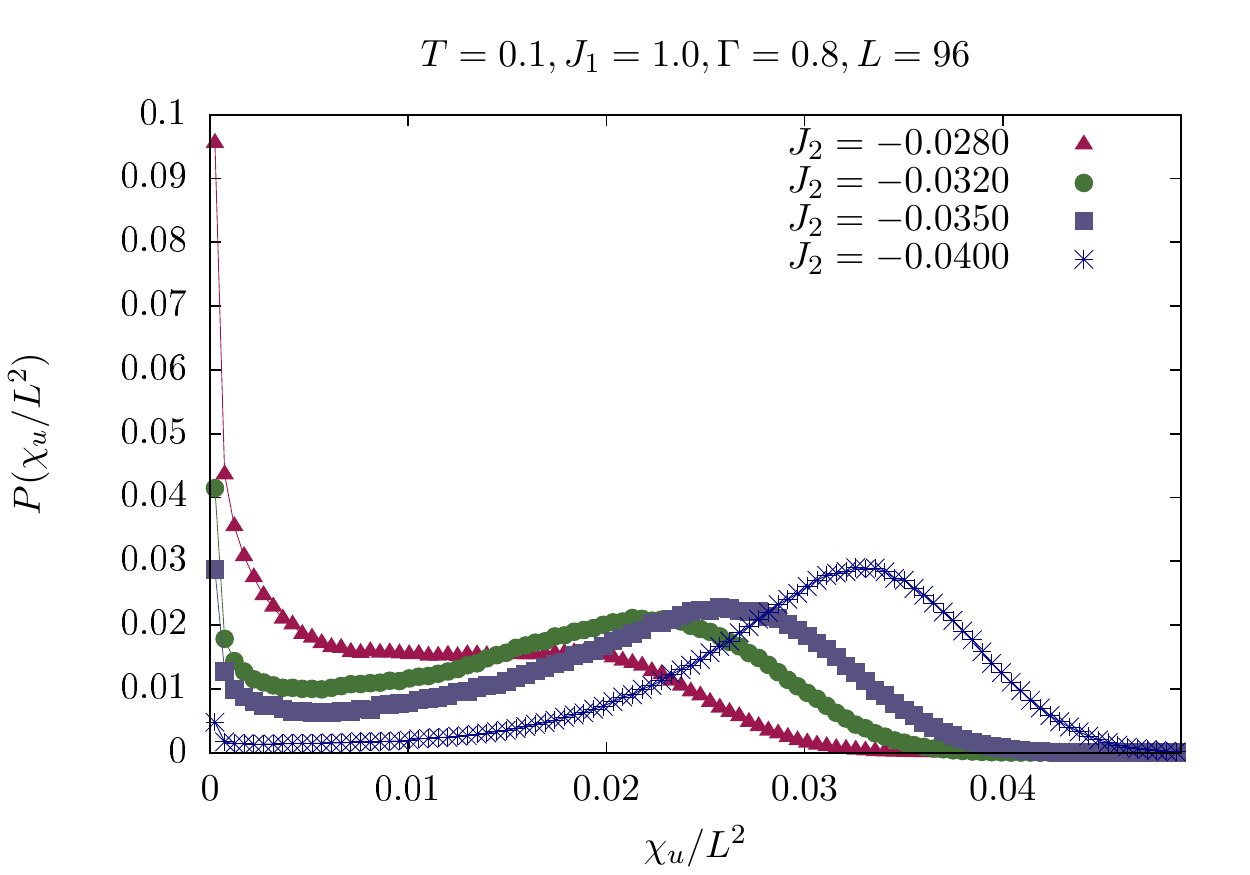}
    \caption{\label{transition} Histograms of $\chi_{u}/L^{2}$ show a characteristic two-peak structure suggestive of a first order transition. All other
temperature and energy scales are measured in units of $J_1$ which is set to unity.} 
  \end{figure}

When ferromagnetic second-neighbour interactions $J_{2}<0$ of sufficient magnitude
are present, one expects the ground state ordering pattern to change to ferrimagnetic
three-sublattice order.\cite{Nienhuis_Hilhorst_Blotte} In recent work,\cite{Biswas_Rakala_Damle}, the threshold value of $J_{2}$ corresponding to this onset of ferrimagnetism was estimated to be roughly $J_{2c} \approx -0.03$. With a view towards comparing the
melting behaviour of this ferrimagnetic three-sublattice order with the two-step
melting of antiferromagnetic three-sublattice order, we also study the effect of
thermal fluctuations at $J_2 =-0.1$, {\em i.e.}  deep in this ferrimagnetic three-sublattice ordered state. We find that long-range order is again lost via a two-step melting process,
with an intermediate power-law ordered phase. The locations of the upper and lower
transitions that demarcate the extent of the power-law ordered phase are obtained
as before. This is displayed in Fig.~\ref{cross2}. 
 Above $T_2$, the static order parameter susceptibility again collapses quite
convincingly on to the Kosterlitz-Thouless finite-size scaling form. This is shown in Fig.~\ref{scaling2}.

With these preliminaries out of the way, we are now in a position to study  in a unified way the behaviour
of the uniform easy-axis susceptibility $\chi_{u}$ in the power-law ordered phase associated with the
two-step melting of antiferromagnetic three-sublattice order as well as ferrimagnetic
three-sublattice order. As mentioned earlier, our goal is
to test a recent prediction\cite{Damle} that $\chi_{u}$ provides a
thermodynamic signature of two step melting due to the presence of a singular
$B$ dependence: $\chi_{u} (B) \sim  |B|^{-\frac{4 - 18 \eta}{4-9\eta}}$ for $\eta(T) \in (1/9,2/9)$.

Here, we test this via the equivalent prediction\cite{Damle} for the finite-size
susceptibility $\chi_{u}(L)$ of an $L \times L$ sample when $B=0$:  $\chi_{u}(L) \sim L^{2-9\eta}$  for  $\eta(T) \in (1/9,2/9)$. In Landau theory terms, this singularity in $\chi_u$ is a direct consequence of a symmetry-allowed coupling of the form $m_{\mathrm{cl}} \lvert \psi_{\mathrm{cl}} \rvert^3 \cos(3\theta)$ between the static component $m_{\mathrm{cl}}(\vec{r})$ of the uniform magnetization density and the order parameter field $\psi_{\mathrm{cl}}$. In the power-law ordered
phase, this coupling is predicted\cite{Damle} to cause $m_{\mathrm{cl}}$ to have the same power-law correlations as $\cos(3 \theta)$, leading to a singular $\chi_{u}$ {\em independent}
of whether the low temperature ordered state is ferrimagnetic or antiferromagnetic.
Thus, while the predicted effect is particularly counter-intuitive for the antiferromagnetic
case, {\em i.e} with $J_2=0$ for the system under consideration, the underlying
mechanism is expected to be the same at $J_2 = -0.1$ as well. 

 As is clear from Fig.~\ref{powerlaw1}, simultaneous fits of  $\chi_{\mathbf{Q}}$ to the form $k_{1}L^{2-\eta}$ and $\chi_{u}$ to the form $k_{2}L^{2-9\eta}$ work rather well at three different
points in the power-law ordered phase associated with the two-step melting of antiferromagnetic three-sublattice order. This can be compared to similar fits in Fig.~\ref{powerlaw2} for
the same quantities  in the power-law ordered phase associated with the two-step
melting of ferrimagnetic three-sublattice order.
As is clear from these results, the uniform susceptibility does indeed provide
a thermodynamic signature of the power-law ordered phase, independent of
the ferri/antiferromagnetic nature of the low-temperature three-sublattice ordered phase,
exactly as predicted by the effective field theoretical arguments of Ref.~\onlinecite{Damle}.

A similar argument, which identifies $\beta^{-1} \int_0^{\beta}\sigma^{x}_{{\mathbf{Q}}}(\tau)$ with $\psi^2_{\mathrm{cl}}$ on symmetry grounds, immediately predicts that $\chi^{xx}_{{\mathbf{Q}}} \sim L^{2-4\eta}$ throughout the power-law ordered phase. As is clear from
Fig.~\ref{transversepowerlaw}, our data for $\chi^{xx}_{{\mathbf{Q}}} $ is seen to be completely consistent
with this prediction as well.

Finally, we comment on the nature of the transition between the antiferromagnetic and ferrimagnetic three sublattice ordered states.  In previous work which studied\cite{Biswas_Rakala_Damle}
relatively small samples at moderately low temperatures in the vicinity of this transition, the phase of the estimator for the three-sublattice order parameter, as measured in the 
Quantum Monte Carlo simulations, was seen to be distributed more or less uniformly in the interval
$(0,2\pi)$. If this behaviour were to persist to larger sizes, it would be indicative
of a power-law ordered phase that interpolates between the antiferromagnetic
and ferrimagnetic three-sublattice ordered phases at nonzero temperature. However, from the Landau
theory considerations of Sec.~\ref{PhasesandTransitions}, we see that the
two generic possibilities for this phase transition are first-order behaviour,
or an intermediate mixed-phase. An intervening power-law ordered phase
can, in this picture, only arise in the fine-tuned limiting case where $\lambda_{12}$
and higher order anisotropies are all absent. With this in mind, we measure the
histogram of the estimator for $\chi_{u}/L^{2}$ to look for signals of phase coexistence
in the transition region.
These histograms are shown in Fig.~\ref{transition}. The two-peak nature
of these histograms suggests that the transition is in fact of a weakly first-order type.
This is consistent with the fact that the $L$-dependence of $\chi_{\mathbf{Q}}$ is certainly
not a power-law, and the fact that Binder ratios of the estimator of $\chi_{\mathbf{Q}}$ also
do not show a clear crossing (indicative of a second-order transition),\cite{Binder_ZPHYSB,Binder_PRL} nor do they stick (as they would in a power-law ordered phase).~\cite{Challa_Landau} However, we do not see any indications
of non-monotonic Binder ratios\cite{Binder_Landau} of the type expected in the vicinity of first order transitions.
Thus, while our data is suggestive of a weakly first-order transition, more work is needed
to clarify the precise nature of this transition.
 
\section{Discussion}
Thus, we have obtained fairly convincing evidence for a 
singular uniform easy-axis susceptibility $\chi_u(B)$ in the power-law ordered
phase associated with the two-step melting of antiferromagnetic three-sublattice order
in triangular lattice transverse-field Ising antiferromagnets. This (at-first-sight) counter-intuitive thermodynamic signature of two-step melting is already of
some general interest, since the transverse-field
Ising antiferromagnet on the triangular lattice is a paradigmatic example of the interplay
between quantum fluctuations and frustrated classical interactions.
Of course, this thermodynamic signature of two-step melting would be of much greater
interest and direct experimental relevance if the model Hamiltonian $H_{\rm Ising}$ were to emerge as
a good description of magnetic exchange interactions in some frustrated magnet.

In this context, it should
be noted that a closely related model Hamiltonian, the one-dimensional transverse
field Ising chain, does serve as a good starting point for the theoretical
description of an interesting quantum phase transition in the magnetic material
Columbite.\cite{Kinross_etal,Lee_etal,Coldea_etal} Columbite can be thought of as a triangular array of one dimensional
chains of magnetic moments, with strong intra-chain coupling between the moments
and weak inter-chain couplings. This hierarchy of exchange-couplings allows
for a theoretical description in terms of the quantum critical properties of the one-dimensional transverse-field Ising chain. It is possible that other materials, with somewhat different exchange pathways but the same strong easy-axis anisotropy, may have much stronger exchange couplings within a triangular plane, and much
weaker couplings between planes. For such a material, the model Hamiltonian $H_{\rm Ising}$ could in the same way serve as a good theoretical description, and our results on this
thermodynamic signature of two-step melting could then be of direct experimental relevance. We hope that our results provide some motivation for exploring this
possibility.

  \section{Acknowledgements}
  \label{Acknowledgements}
  Our computational work was made possible by the computational resources
  of the Department of Theoretical Physics of the Tata Institute of Fundamental
  Research, as well as by computational resources funded by 
  DST (India) grant DST-SR/S2/RJN-25/2006. The analysis of our Monte Carlo
data was greatly facilitated by the general-purpose file-handling and data-analysis scripts
developed by Geet Rakala.
  \bibliography{mybib}

  \end{document}